\documentclass[11pt]{article}
\usepackage{blois,epsfig}

\def\be{\begin{equation}}
\def\ee{\end{equation}}
\def\bea{\begin{eqnarray}}
\def\eea{\end{eqnarray}}

\def\adeg{^{\circ}}

\def\amin{^\prime}
\def\Msol{\mbox{ }M_{\odot}}
\def\Rsol{\mbox{ }R_{\odot}}

\begin{document}
\vspace*{4cm}
\title{FERMI-GST: A NEW VIEW OF THE $\gamma$-RAY SKY}

\author{ S. CHATY on behalf of the {\it Fermi}-LAT collaboration }

\address{Laboratoire AIM (UMR 7158 CEA/DSM-CNRS-Universit\'e Paris Diderot),
Irfu/Service d'Astrophysique, CEA-Saclay,
FR-91191 Gif-sur-Yvette Cedex, France, e-mail: chaty@cea.fr
}

\maketitle\abstracts{
The Large Area Telescope on the {\it Fermi} $\gamma$-ray Space Telescope (FGST, ex-GLAST) provides unprecedented sensitivity for all-sky monitoring of $\gamma$-ray activity. It is an adequate telescope to detect transient sources, since the observatory scans the entire sky every three hours and allows a general search for flaring activity on daily timescale. This search is conducted automatically as part of the ground processing of the data and allows a fast response --less than a day-- to transient events. Follow-up observations in X-rays, optical, and radio are then performed to attempt to identify the origin of the emission and probe the possible existence of new transient $\gamma$-ray sources in the Galaxy.
Since its launch on 11th June 2008, {\it Fermi}-LAT has detected nearly 1500 $\gamma$-ray sources, nearly half of them being extragalactic. After a brief census of detected celestial objects, we report here on the LAT results focusing on Galactic transient binary systems. The {\it Fermi}-LAT has detected 2 $\gamma$-ray binaries, a microquasar and an unexpected new type of $\gamma$-ray source: a symbiotic nova.
}

\section{Introduction}\label{introduction}
There has been two firmly established classes of variable sources in the high-energy $\gamma$-ray sky: blazars and $\gamma$-ray bursts (GRBs). The Energetic $\gamma$-Ray Experiment Telescope ({\it EGRET}) on the Compton $\Gamma$-Ray Observatory discovered a population of variable $\gamma$-ray blazars above 100 MeV \cite{hartman:1999}. However, {\it EGRET} also left the legacy of a large fraction of unidentified sources in the 3EG catalog.
Many of these were found at low Galactic latitudes and
believed to be Galactic in nature. {\it EGRET} established
$\gamma$-ray pulsars as a Galactic population and these
were thought to contribute to the unidentified sources.
Several studies found indications of variability in some
of the sources along the Galactic Plane (GP) \cite{torres:2001,nolan:2003,wallace:2000}, a behavior not expected of the pulsars, which are steady
on these timescales. Additionally, no blazar counterpart
was identified in several cases. This suggested the
possible existence of a new Galactic $\gamma$-ray class.

Several sources showed both strong variability and a
convincing lack of a blazar counterpart within the {\it EGRET} localization
errors, like for instance 3EG\,J0241+6103, 3EG\,J1824-1514 and 3EG\,J1837-0423 (GRO\,J1834-04). The most dynamic example is GRO\,J1834-04, which produced an intense outburst in June 1995 \cite{tavani:1997}. The flux above 100\,MeV in a 3.5 day period was found to be a factor of 7 brighter than in later observations of the region. Notably, no blazar counterpart is known within the 99\% {\it EGRET} error contour. The absence of a flat spectrum radio source at the levels typical of the {\it EGRET} blazars made this a candidate for a different type of $\gamma$-ray emitter. The proximity of such a unique outburst to the inner Galaxy led to speculation of a possible Galactic origin. The question of the progenitor of this activity remains as well as the broader question of the existence of similar sources of this type. As stated in Ref.~ \cite{tavani:1997}, ``other unidentified {\it EGRET} sources near the GP appear to be time variable with V$>1.5$ in Ref.~ \cite{mclaughlin:1996ApJ473.763}''.

\section{The {\it Fermi}-LAT $\gamma$-ray sky}

After a bit more than 2 years of observations, the {\it Fermi}-LAT $\gamma$-ray sky is already a pretty rich one. It includes $\sim$1500 sources, among which we find:

\begin{itemize}

\item the Sun (Inverse Compton interactions of the solar radiation field and proton interactions in the outer atmosphere), the Moon (from proton interactions in the rock),

\item $\sim$5 Supernova remnants (SNRs) and Pulsar Wind Nebulae (PWN) ($\gamma$-ray emission produced by cosmic ray protons deflected by magnetic field, and accelerated in SNRs, {\it mechanism predicted by Enrico Fermi in 1949}),

\item 64 pulsars and millisecond pulsars (16 discovered by LAT, among which 12 brand new $\gamma$-ray only pulsars), 

\item 8 globular clusters (magnetospheric emission, allowing to study the millisecond pulsar population spun up by mass accretion from binary companions), 

\item 2 $\gamma$-ray binaries, 1 microquasar (X-ray binary) and 1 symbiotic nova,

\item $\sim$750 Active Galactic Nuclei and blazars, starburst galaxies, the Small Magellanic Cloud, the Large Magellanic Cloud, 

\item Galactic diffuse emission, extragalactic $\gamma$-ray background (for which AGNs contribute only to $\sim$30\%, the rest being probably particle acceleration in normal star-forming galaxies, cluster of galaxies, and cosmological dark matter annihilation), 

\item 16 GRBs (GRB090510, at a distance of 7.3 Gyr, exhibited the fastest ejected matter, at 0.9999995c, and allowed a quantum gravity test, showing that there was no strong energy dependance in $c$, a 31 GeV photon arriving only 0.83s after the 1st X-ray photon),

\item and still $\sim$500 unidentified yet sources...

\end{itemize}

Therefore not only the LAT allows to detect astrophysical sources, it also allows us to perform cosmology, particle astrophysics studies, measure of the cosmic-ray specrum, dark-matter annihilation, etc...

\section{LAT Detection of $\gamma$-ray transients} \label{detection}

The {\it Fermi} Large Area Telescope (LAT) is very well suited
for monitoring variability in the GeV sky. The
large effective area ($>8000$\,cm$^2$ on axis above 1\,GeV),
wide field-of-view ($\sim 2.4$\,sr), and excellent angular resolution
(better than $1\adeg$ above 1\,GeV) greatly enhance
the sensitivity to transient activity in comparison to
previous $\gamma$-ray instruments in this energy range.
A notable departure from previous observations is that
the energy range of the LAT (20\,MeV to $>300$\,GeV)
extends above that covered by {\it EGRET}. The combination
of the wide field-of-view with the sky scanning observational
mode supplies coverage of the sky every $\sim 3$\,hours
(2 orbits). This enables the detection of fainter objects
in shorter intervals than previously possible. 
Also, the localizations for sources above threshold are much better than $1\adeg$ even on short timescales.
%

All these capabilities are critical for triggering rapid multiwavelength follow-up observations of LAT transients. The LAT has so far detected three transient events near the GP, high-confidence detections that appeared in multiple 6 hour and daily ASP searches, that have not been associated with blazars or other known sources: 3EG\,J0903-3531, Fermi\,J0910-5041 and Fermi\,J1057-6027. The $\gamma$-ray characteristics and the multiwavelength follow-up observations that were triggered shortly after their detections have been intensively described in Chaty et al. (2010) \cite{chaty:2010}. Here we now report on identified Galactic sources.

\section{Galactic transient sources}

\subsection{Two $\gamma$-ray binaries: LSI\,$+61 \adeg 303$ \& LS\,$5039$}

\subsubsection{LSI\,$+61 \adeg 303$}

One of the transient {\it EGRET} sources has emerged as a new type of $\gamma$-ray source with the rediscovery of 3EG\,J0241+6103 (COS-B 2CG\,135+01), which was associated, although the position was uncertain, with LSI\,$+61 \adeg 303$, a "prominent radio flaring star system'' \cite{tavani:1998}, high mass X-ray binary (HMXB) system located at 2 kpc and constituted of a neutron star orbiting a B0\,Ve star on an excentric orbit, with a 26.5\,days period. A daily/monthly variability was seen with {\it EGRET}, but no periodicity detected \cite{tavani:1998}. A TeV source has then been detected by MAGIC and then VERITAS, at this position, with a periodic signal modulated at the orbital period. {\it Fermi} has detected a $\gamma$-ray source, 0FGL\,J0240.3+6113, at the position RA=40.076, DEC=61.233 with a 95\% error radius of $1.8 \amin$, consistent with the optical counterpart, and with a periodicity of $26.6 \pm 0.5$\,days, the emission peaking at the periastron, and a spectrum reminiscent of the pulsars, therefore definitely identifying in the MeV-GeV domain the first $\gamma$-ray binary source with the HMXB system, the pulsar wind interacting with the companion stellar wind \cite{abdo:2009ApJ701L123}.

\subsubsection{LS\,$5039$}

The second interesting source is the case of 3EG\,J1824-1514, detected by {\it EGRET}, but without modulation, as spatially coincident with LS\,5039, an HMXB constituted of a likely neutron star orbiting around an O6.5 star. HESS detected a periodic signal modulated at the orbital period of 3.91 days. {\it Fermi} firmly detected at more than $12 \sigma$ a periodic source, modulated at $3.91 \pm 0.05$ days, in a complicated region with a very intense Galactic diffuse emission. The second $\gamma$-ray binary was definitely detected by {\it Fermi} \cite{abdo:2009ApJ706L56A}.

\subsection{A microquasar: Cyg X-3}

Recently, {\it Fermi} has detected a variable $\gamma$-ray source coinciding with the position of the X-ray binary and microquasar Cygnus X-3, modulated at its short orbital period of 4.8 hours. Cygnus X-3 is an HMXB system located at a distance of $\sim 7$\,kpc, with a compact object of nature still matter in the debate, orbiting a Wolf-Rayet (WR) star \cite{abdo:2009Science326.1512}.
The high-energy source is detected during periods of relativistic ejection events. The emission is likely coming from the interaction between the wind of the WR star and the accretion disk, the corona, and probably also the relativistic jets... Cygnus X-3 is a unique source among X-ray binaries: it has a very tight orbit, a strong WR wind ($10^{-5} \Msol$/yr, $10^3$\,km/s, high density), favorizing the UV stellar photons to strike ultrarelativistic particles (e$^-$), creating $\gamma$-rays. This source is therefore a good candidate for neutrino and cosmic-ray protons emitter.

\subsection{A symbiotic nova: V407 Cyg}

Finally, one of the most recent results is that {\it Fermi} has unexpectedly detected $\gamma$-ray emission for 15 days from a nova in the symbiotic binary V407 Cygni, containing a white dwarf and a red giant star about $500 \Rsol$ \cite{abdo:2010Science}. 
The mechanism is the following one:
for decades to centuries, the white dwarf captures some of the red giant's wind, which accumulates on its surface, up to the point when this gas produces a thermonuclear explosion: the ``nova'' outburst occurs. The shock wave created by this blast, composed of high-speed particles, ionized gas and magnetic fields, is accelerated from $\sim 0.01c$ to $c$, and then enters in collision with the red giant wind, creating the $\gamma$-ray emission.

\section{Conclusion}

The {\it Fermi}-LAT $\gamma$-ray sky is therefore very rich, with many astrophysical sources of various types detected. And it only observed for a bit more than 2 years, so many more sources are expected!
Apart from the classical $\gamma$-ray sources --blazars and pulsars--, the LAT has detected plenty of SNRs, PWN, globular clusters, and even GRBs. It has also detected 2 $\gamma$-ray binaries, 1 microquasar, and a previously unexpected symbiotic nova...

The observatory is constantly watching for new sources, especially transient sources. Furthermore, the yet unidentified sources are not to be forgotten. The task of identifying the counterparts for unidentified transients remains challenging though. While greatly reduced from {\it EGRET}, the LAT error circles remain large compared to the resolution of telescopes in other wavebands. Additional arguments based on temporal and spectral characteristics are required to support firm associations. Ultimately, identifications of the Galactic transients require observations of related variability between the $\gamma$-ray source and a candidate Galactic counterpart at lower frequency. In the absence of a detection of significant activity of the potential radio and X-ray counterparts for the LAT transients, they remain unidentified. These sources continue to be monitored regularly for $\gamma$-ray activity as a part of the {\it Fermi} sky survey observations.

\section*{Acknowledgments}
The {\it Fermi}-LAT Collaboration acknowledges support
from a number of agencies and institutes for both the
development and operation of the LAT as well as the scientific data analysis. These include NASA and the DOE
in the United States, the CEA/Irfu and IN2P3/CNRS
in France, ASI and INFN in Italy, MEXT, KEK, and
JAXA in Japan, and the K.A. Wallenberg Foundation,
the Swedish Research Council and the National Space
Board in Sweden. Additional support from INAF in Italy
for science analysis during the operations phase is also
gratefully acknowledged.
This work was supported by the Centre National d'Etudes Spatiales (CNES), based on observations obtained through MINE: the Multi-wavelength {\it INTEGRAL}/{\it Fermi} NEtwork.

\section*{References}

\end{document}